# Predicting potential SARS-CoV-2 spillover and spillback in animals

Zi Hian TAN[a], Kian Yan YONG[a] and Jian-Jun SHU[a,✉]

**ABSTRACT**

***Background***: The COVID-19 pandemic is spreading rapidly around the world, causing countries to impose lockdowns and efforts to develop vaccines on a global scale. However, human-to-animal and animal-to-human transmission cannot be ignored, as severe acute respiratory syndrome coronavirus 2 (SARS-CoV-2) can spread rapidly in farmed and wild animals. This could create a worrying cycle of SARS-CoV-2 spillover from humans to animals and spillback of new strains back into humans, rendering vaccines ineffective.

***Method***: This study provides a key indicator of animals that may be potential susceptible hosts for SARS-CoV-2 and coronavirus infections by analysing the phylogenetic distance between host angiotensin-converting enzyme 2 and the coronavirus spike protein. Crucially, our analysis identifies animals that are at elevated risk from a spillover and spillback incident.

***Results***: One group of animals has been identified as potentially susceptible to SARS-CoV-2 by harbouring a parasitic coronavirus spike protein similar to the SARS-CoV-2 spike protein. These animals may serve as amplification hosts in spillover events from zoonotic reservoirs. This group consists of a mixture of animals infected internally and naturally: minks, dogs, cats, tigers. Additionally, no internal or natural infections have been found in masked palm civet.

***Conclusion***: Tracing interspecies transmission in multi-host environments based solely on *in vitro* and *in vivo* examinations of animal susceptibility or serology is a time-consuming task. This approach allows rapid identification of high-risk animals to prioritize research and assessment of the risk of zoonotic disease transmission in the environment. It is a tool to rapidly identify zoonotic species that may cause outbreaks or participate in expansion cycles of coexistence with their hosts. This prevents the spread of coronavirus infections between species, preventing spillover and spillback incidents from occurring.

**KEYWORDS** SARS-CoV-2; outbreak; human-to-animal and animal-to-human transmissions; spillover and spillback; infectious disease

# Introduction

Since the outbreak of the novel coronavirus caused by the 2019 novel coronavirus (2019-nCoV)[1,2] at the end of 2019, the virus was subsequently identified as severe acute respiratory syndrome coronavirus 2 (SARS-CoV-2) and has spread rapidly throughout the country world. Since the outbreak of severe acute respiratory syndrome (SARS), it has been known that animals can become hosts for amplification by adding mutations, thereby generating new strains. For example, there was an outbreak at a mink farm that caused a mutation in SARS-CoV-2[3]. Although this mutation did not lead to new strains of concern, it highlights the importance of managing spillover from human to animals to prevent spillback transfer. Currently, the susceptibility of common domestic and farmed animals to SARS-CoV-2 is established; however, there are more contacts between humans and animals than currently assessed, and a single spillover incident of an amplification host may lead to the creation of new SARS-CoV-2 strains.

[a] School of Mechanical & Aerospace Engineering, Nanyang Technological University, 50 Nanyang Avenue, Singapore 639798.
✉ Correspondence should be addressed to Jian-Jun SHU. *E-mail address*: mjjshu@ntu.edu.sg.

In order to identify susceptible animals, the main determinants of infection must first be studied. SARS-CoV-2 shares the same cellular receptor as severe acute respiratory syndrome coronavirus (SARS-CoV), but the spike protein has a higher binding affinity[4,5] to human angiotensin-converting enzyme 2 (hACE2), resulting in higher infectivity. Animal susceptibility has been shown to be consistent with phylogenetic grouping[6,7]; however, most phylogenetic studies of susceptibility have been qualitative. To predict the susceptibility of animals, the phylogenetic distance of angiotensin-converting enzyme 2 (ACE2) in each animal is determined with reference to hACE2. The distance of the coronavirus spike protein from each animal host to the first SARS-CoV-2 sequence is determined and compared together with the ACE2 distance. Our results reveal that the ACE2 of the suspected intermediate animal pangolin is more similar to humans than to bats, and the spike protein of SARS-CoV-2-related pangolin coronavirus (Pangolin-CoV) exhibits more similarity to SARS-CoV-2 than the closest bat coronavirus. Other animals have also been determined to be susceptible to SARS-CoV-2, which provides a framework for further susceptibility testing of animals to prevent spillover and spillback incidents.

## Method

### Sequence

ACE2 amino acid sequence is obtained from UniProt (https://www.uniprot.org/) and the National Center for Biotechnology Information (NCBI) (https://www.ncbi.nlm.nih.gov/). Table 1 lists the 225 ACE2 amino acid sequences.

**Table 1. List of ACE2 amino acid sequences**

| Protein ID | Scientific name | Common name |
|---|---|---|
| Q9BYF1 | *Homo sapiens* | Human |
| Q8R0I0, Q3URC9 | *Mus musculus* | Mouse |
| Q5EGZ1, D3ZYK4, C7ECU5, A0A0G2JXU8 | *Rattus norvegicus* | Rat |
| Q56NL1 | *Paguma larvata* | Masked palm civet |
| Q5RFN1, H2PUZ5, NP_001124604.1 | *Pongo abelii* | Sumatran orangutan (*Pongo pygmaeus abelii*) |
| Q56H28, A0A5F5XDN9 | *Felis catus* | Domestic cat (*Felis silvestris catus*) |
| Q58DD0, A0A452DJE0, Q2HJI5 | *Bos taurus* | Bovine |
| G1RE79 | *Nomascus leucogenys* | Northern white-cheeked gibbon (*Hylobates leucogenys*) |
| I3M887 | *Ictidomys tridecemlineatus* | Thirteen-lined ground squirrel (*Spermophilus tridecemlineatus*) |
| G3QWX4 | *Gorilla gorilla gorilla* | Western lowland gorilla |
| A0A0D9RQZ0 | *Chlorocebus sabaeus* | Green monkey |
| | | (*Cercopithecus sabaeus*) |
| Q2WG88 | *Mustela putorius furo* | Ferret (*Mustela furo*) |
| A0A2K5X283 | *Macaca fascicularis* | Crab-eating macaque (Cynomolgus monkey) |
| J9P7Y2, F1P7C5, A0A5F4BS93, A0A5F4CXG9 | *Canis lupus familiaris* | Domestic dog (*Canis familiaris*) |
| H0VSF6 | *Cavia porcellus* | Guinea pig |
| G1PXH7 | *Myotis lucifugus* | Little brown bat |
| A0A2R9BKD8, A0A2R9BJK0 | *Pan paniscus* | Pygmy chimpanzee (Bonobo) |
| F6V9L3 | *Equus caballus* | Horse |
| A0A096N4X9 | *Papio anubis* | Olive baboon |
| W5PSB6 | *Ovis aries* | Sheep |
| F6WXR7 | *Monodelphis domestica* | Gray short-tailed opossum |
| G3T6Q2 | *Loxodonta africana* | African bush elephant |
| A0A2K6NFG7 | *Rhinopithecus roxellana* | Golden snub-nosed monkey (*Pygathrix roxellana*) |
| A0A2K5DQI6 | *Aotus nancymaae* | Ma's night monkey |
| A0A0N8EUX7 | *Heterocephalus glaber* | Naked mole-rat |
| A0A2K6D1N8 | *Macaca nemestrina* | Pig-tailed macaque |
| F7AH40, B6DUG6, B6DUE3, B6DUE5, B6DUF5, B6DUE6, B6DUF2, B6DUE2, B6DUG5, B6DUE4, B6DUF6, B6DUF7, B6DUE9, B6DUE8, B6DUG2, B6DUG0, B6DUF1, B6DUF9, B6DUE1, B6DUE7, B6DUF3, B6DUF8 | *Macaca mulatta* | Rhesus macaque |
| A0A2K5ZV99 | *Mandrillus leucophaeus* | Drill (*Papio leucophaeus*) |
| A0A2K5KSD8 | *Cercocebus atys* | Sooty mangabey (*Cercocebus torquatus atys*) |
| A0A452EVJ5, A0A452EVU0, A0A452EVM2 | *Capra hircus* | Goat |
| A0A1U7TY97 | *Tarsius syrichta* | Philippine tarsier |
| A0A2Y9M9H3 | *Delphinapterus leucas* | Beluga whale |
| K7FJ41 | *Pelodiscus sinensis* | Chinese softshell turtle (*Trionyx sinensis*) |
| A0A2K6GHW5 | *Propithecus coquereli* | Coquerel's sifaka (*Propithecus verreauxi coquereli*) |
| A0A452TT30, A0A384CIJ9, A0A452TT37, A0A452TTE2, A0A452TT60, A0A452TT98, A0A452TTD2, A0A452TTF7, A0A452TTE1 | *Ursus maritimus* | Polar bear (*Thalarctos maritimus*) |
| F1NHR4, A0A5J6CU64 | *Gallus gallus* | Red junglefowl |
| U3JP73 | *Ficedula albicollis* | Collared flycatcher (*Muscicapa albicollis*) |
| A0A5F4W5D9, F7CNJ6 | *Callithrix jacchus* | White-tufted-ear marmoset |
| H0WMI5 | *Otolemur garnettii* | Small-eared galago (Garnett's greater bushbaby) |
| F7ABF9, F6PSC4, F6PSI0, F6PU11, A0A5S6MIJ1, A0A6I8QCQ6, A0A6I8SUJ0, A0A6I8S716, A0A6I8RH75 | *Xenopus tropicalis* | Western clawed frog (*Silurana tropicalis*) |
| F7FDA2 | *Ornithorhynchus anatinus* | Duckbill platypus |
| A0A2K6SBD4 | *Saimiri boliviensis boliviensis* | Bolivian squirrel monkey |
| G1MC42 | *Ailuropoda melanoleuca* | Giant panda |
| A0A1U7QTA1 | *Mesocricetus auratus* | Golden hamster |
| G1NPB8, G5E7W8 | *Meleagris gallopavo* | Wild turkey |
| A0A2K5PYM0 | *Cebus capucinus imitator* | Panamanian white-faced capuchin |
| G1KTF3 | *Anolis carolinensis* | Green anole (American chameleon) |
| A0A3Q7RAT9 | *Vulpes vulpes* | Red fox |
| A0A2Y9S5T9 | *Physeter macrocephalus* | Sperm whale (*Physeter catodon*) |
| A0A452R1Z9 | *Ursus americanus* | American black bear (*Euarctos americanus*) |
| A0A1S3GHT7, A0A1S3GFD6 | *Dipodomys ordii* | Ord's kangaroo rat |
| A0A4W2H6E0, A0A4W2H3A1 | *Bos indicus x Bos taurus* | Hybrid cattle |
| A0A4X2M679 | *Vombatus ursinus* | Common wombat |
| A0A2K6LKA0 | *Rhinopithecus bieti* | Black snub-nosed monkey (*Pygathrix bieti*) |
| A0A2J8KU96, A0A2I3S8E3 | *Pan troglodytes* | Chimpanzee |
| K7GLM4 | *Sus scrofa* | Wild boar |
| A0A2K5JE65 | *Colobus angolensis palliatus* | Peters' Angolan colobus |
| U3J4G2 | *Anas platyrhynchos platyrhynchos* | Northern mallard |
| A0A1L8HCX9 | *Xenopus laevis* | African clawed frog |
| H3B2W0 | *Latimeria chalumnae* | Coelacanth |
| I3J601, A0A669DS63, A0A669B8J3, A0A669BJI9, A0A669D2Q6, A0A669F4X0 | *Oreochromis niloticus* | Nile tilapia (*Tilapia nilotica*) |
| A0A1S3APE5 | *Erinaceus europaeus* | Western European hedgehog |
| A0A2D0Q7Z4 | *Ictalurus punctatus* | Channel catfish (*Silurus punctatus*) |
| A0A340Y3Y6 | *Lipotes vexillifer* | Yangtze river dolphin |
| E7F9E5, Q5U380 | *Danio rerio* | Zebrafish (*Brachydanio rerio*) |
| A0A2U3X0M3 | *Odobenus rosmarus divergens* | Pacific walrus |
| A0A3Q0H852, A0A3Q0H3J6 | *Alligator sinensis* | Chinese alligator |
| A0A452CBT6 | *Balaenoptera acutorostrata scammoni* | North Pacific minke whale (*Balaenoptera davidsoni*) |
| A0A3Q7TE16 | *Ursus arctos horribilis* | Grizzly bear |
| Q1LZX8 | *Chlorocebus aethiops* | Green monkey (*Cercopithecus aethiops*) |
| E2DHI3, ADN93471.1 | *Rhinolophus macrotis* | Big-eared horseshoe bat |
| A4PIG8, D8WU01 | *Rousettus leschenaultii* | Leschenault's rousette |
| Q2PGE1 | *Procyon lotor* | Raccoon |
| E2DHI4, E2DHI7, U5WHY8 | *Rhinolophus sinicus* | Chinese rufous horseshoe bat |
| E2DHI9 | *Rhinolophus pusillus* | Least horseshoe bat |
| E2DHI2, B6ZGN7, A0A671F9Q9, A0A671F0T6 | *Rhinolophus ferrumequinum* | Greater horseshoe bat |
| B4XEP4 | *Nyctereutes procyonoides* | Raccoon dog (*Canis procyonoides*) |
| A0A3Q7N3M7 | *Callorhinus ursinus* | Northern fur seal |
| A0A341BCI8 | *Neophocaena asiaeorientalis asiaeorientalis* | Yangtze finless porpoise (*Neophocaena phocaenoides asiaeorientalis*) |
| A0A2Y9GBR2, A0A2Y9GEI9 | *Neomonachus schauinslandi* | Hawaiian monk seal (*Monachus schauinslandi*) |
| A0A4W4EE33, A0A4W4EFY7 | *Electrophorus electricus* | Electric eel (*Gymnotus electricus*) |
| A0A663EPL4 | *Aquila chrysaetos chrysaetos* | Golden eagle |
| A0A5P9VP25 | *Tadorna cana* | South African shelduck |
| A0A667IF49 | *Lynx canadensis* | Canada lynx |
| A0A220QT48 | *Sus scrofa domesticus* | Domestic pig |
| A0A673UPR4 | *Suricata suricatta* | Meerkat |
| A0A663M979, A0A663M8Y2, A0A663M7K6, A0A663M8A8 | *Athene cunicularia* | Burrowing owl (*Speotyto cunicularia*) |
| A0A6C0PIH2 | *Oryctolagus cuniculus* | European rabbit |
| A0A151N089 | *Alligator mississippiensis* | American alligator |
| A0A1A8AXC5 | *Nothobranchius furzeri* | Turquoise killifish |
| H0ZCK6, H0ZYW8, A0A674GKE4, A0A674GHV0, A0A674GDZ7, A0A674GJP6 | *Taeniopygia guttata* | Zebra finch (*Poephila guttata*) |
| A0A1V4JC49 | *Patagioenas fasciata monilis* | Band-tailed pigeon |
| A0A665VWQ8, A0A665VWR3 | *Echeneis naucrates* | Live sharksucker |
| A0A4Z2GEX4 | *Liparis tanakae* | Tanaka's snailfish |
| A0A669PPG5, A0A669PSZ2, A0A669Q5K7, A0A669PKV0, A0A669Q5M4 | *Phasianus colchicus* | Common pheasant |
| A0A671T498 | *Sinocyclocheilus anshuiensis* | Anshui golden thread catfish |
| A0A4W3HYJ6, A0A4W3I547, A0A4W3I1M1, A0A4W3HYM0, A0A4W3I1M6, A0A4W3IPJ3, A0A4W3HYL1, A0A4W3IPI8, A0A4W3HYL6 | *Callorhinchus milii* | Ghost shark |
| A0A667X0J3 | *Myripristis murdjan* | Pinecone soldierfish |
| A0A218UNR1 | *Lonchura striata domestica* | Bengalese finch |
| A0A2I0MLI2 | *Columba livia* | Rock dove |
| A0A672V5V3 | *Strigops habroptila* | Kakapo |
| A0A1S3SF35 | *Salmo salar* | Atlantic salmon |
| A0A2I4D9L3 | *Austrofundulus limnaeus* | Limnaeus Killifish |

| V8NIH2 | *Ophiophagus hannah* | King cobra (*Naja hannah*) |
|---|---|---|
| A0A0S7FTS4 | *Poeciliopsis prolifica* | Blackstripe livebearer |
| A0A670YAG2 | *Pseudonaja textilis* | Eastern brown snake |
| A0A2U4AJL3 | *Tursiops truncatus* | Atlantic bottle-nosed dolphin (*Delphinus truncatus*) |
| A0A0K8U7D5 | *Bactrocera latifrons* | Malaysian fruit fly (*Chaetodacus latifrons*) |
| A0A2S2P0H8 | *Schizaphis graminum* | Greenbug aphid |
| A0A2H8TEU2 | *Melanaphis sacchari* | Sugarcane aphid |
| QLF98521.1 | *Melogale moschata* | Chinese ferret-badger |
| QLF98526.1 | *Arctonyx collaris* | Hog badger |
| QLF98520.1 | *Tadarida brasiliensis* | Mexican free-tailed bat |
| ACT66266.1 | *Pipistrellus abramus* | Japanese house bat |
| XP_017505752.1 | *Manis javanica* | Pangolin |
| QLH93383.1 | *Manis pentadactyla* | Chinese pangolin |
| QNC68911.1 | *Mustela lutreola biedermanni* | European mink |
| ACT66274.1 | *Phodopus campbelli* | Campbell's dwarf hamster |
| ACT66267.1 | *Chinchilla lanigera* | Long-tailed chinchilla |
| QKE49997.1 | *Cynopterus sphinx* | Greater short-nosed fruit bat |
| QKE49998.1 | *Megaderma lyra* | Greater false vampire bat |
| QNV47311.1 | *Tupaia glis* | Common treeshrew |
| XP_006164754.1 | *Tupaia belangeri chinensis* | Chinese treeshrew |
| XP_032187679.1 | *Mustela erminea* | Short-tailed weasel |
| XP_031301717.1 | *Camelus dromedarius* | Arabian camel |
| ABU54053.1 | *Rhinolophus pearsonii* | Pearson's horseshoe bat |
| XP_011361275.1 | *Pteropus vampyrus* | Large flying fox |
| XP_020768965.1 | *Odocoileus virginianus* | White-tailed deer |
| XP_007090142.2 | *Panthera tigris* | Tiger |

The coronavirus spike protein amino acid sequence of the animal host and the complete coding sequence (CDS) of SARS-CoV from *Paguma larvata* (AY515512.1) are downloaded from the NCBI virus database (https://www.ncbi.nlm.nih.gov/labs/virus/vssi/#/). Only complete sequences are downloaded. Further processing is performed to remove experimental recombinant viruses. The only human coronaviruses used are SARS-CoV-2 (YP_009724390.1) and SARS-CoV (YP_009825051.1), which are used as human coronavirus references.

**Alignment and phylogenetic tree generation**

Since the complete CDS of SARS-CoV from *Paguma larvata* (AY515512.1) does not contain an annotation for the spike protein, this sequence is aligned with the SARS Tor2 spike protein (NC_004718.3:21492-25259) to obtain the spike protein sequence. Sequences are translated into amino acid sequences using MUSCLE[8]. ACE2 alignment[9–13] is performed by Nextflow[14] using MAFFT[15], and phylogenetic trees are generated using IQ-TREE[16] with ultrafast bootstrap[17] and ModelFinder Plus[18]. Both ACE2 and spike protein phylogenetic trees are generated using JTT+F+R6 and WAG+F+R10 models of ACE2 and spike protein sequences, respectively, and selected according to the Bayesian information criterion of ModelFinder Plus.

**Obtain phylogenetic distance**

Phylogenetic distances are obtained from tree files using the Python package TreeSwift[19]. All ACE2 distances are measured from node to hACE2 node (UniProtKB-Q9BYF1). Zoonotic coronavirus spike protein distances are measured from node SARS-CoV-2 (YP_009724390.1) and SARS-CoV (YP_009825051.1) spike proteins. For ACE2 sequences, intraspecific phylogenetic distances are determined to be close (Figure 1); therefore, each animal is represented as an average.

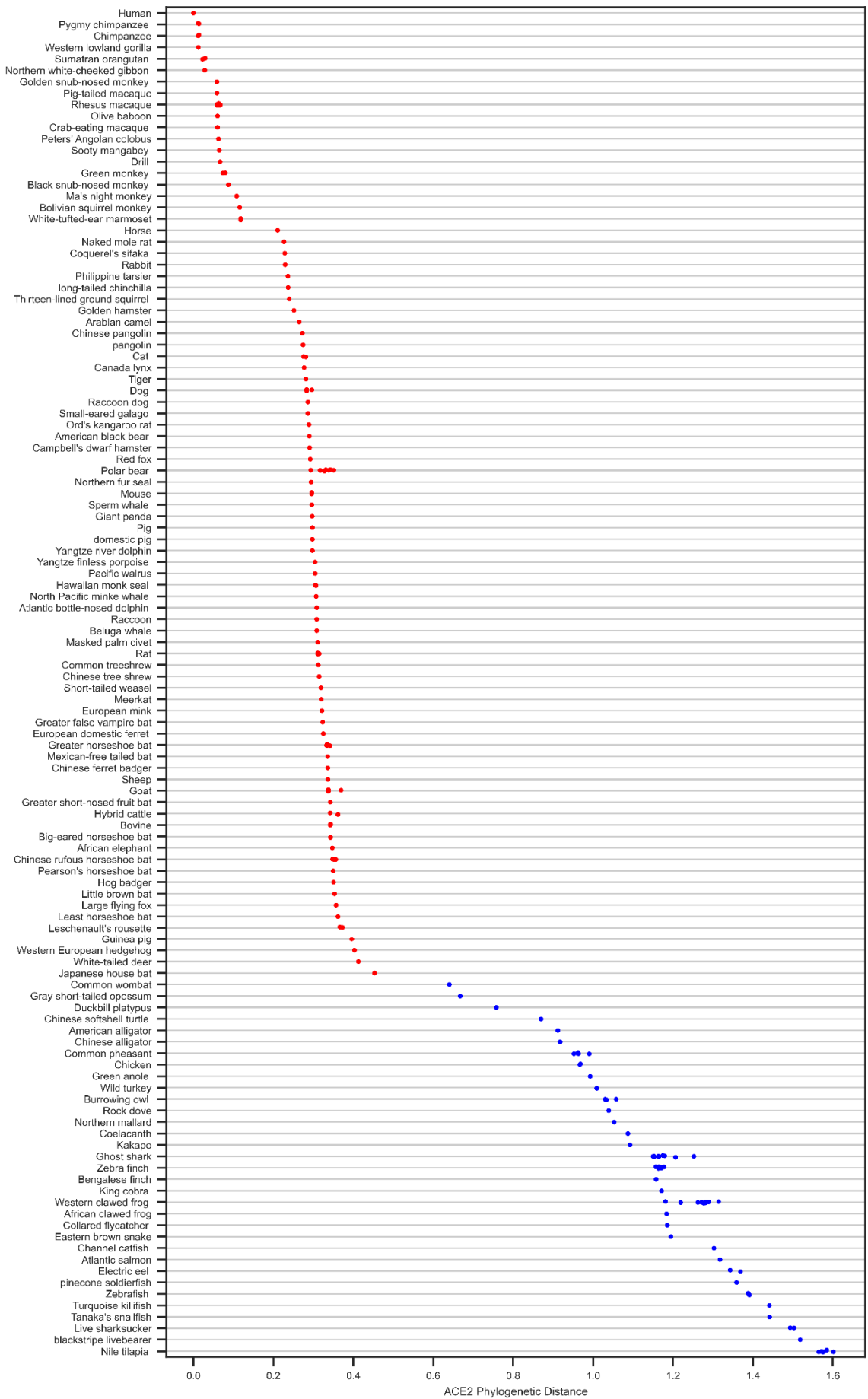

**Figure 1. Distribution of phylogenetic distance of animal ACE2 relative to hACE2.** Only animals with mulitple ACE2 sequences are shown and grouped according to phylogenetic distance from hACE2: close to humans (red < 0.5), distant from humans (0.5 ≤ blue ≤ 2). ACE2 is extremely narrowly distributed in each animal.

In order to compare the phylogenetic distance between ACE2 and spike protein, each animal must have its own ACE2 and parasitic coronavirus spike protein sequences (Table 2). Several subspecies are chosen as representatives of the species. From the coronavirus spike protein sequence, *Pan troglodytes verus* stands for *Pan troglodytes*. From the ACE2 sequence, *Mustela lutreola biedermanni* represents *Mustela lutreola*, and *Lonchura striata domestica* represents *Lonchura striata*.

**Table 2. List of coronavirus spike protein sequences**

| Protein ID | Host | Common name | Species |
|---|---|---|---|
| AWW13519.1 | *Pan troglodytes* | Chimpanzee | *Betacoronavirus 1* |
| ADC35511.1 | *Chlorocebus aethiops* | Green monkey (*Cercopithecus aethiops*) | *Severe acute respiratory syndrome–related coronavirus (SARSr-CoV)* |
| BAS18866.1 | *Equus caballus* | Horse | *Betacoronavirus 1* |
| AFE48795.1 | *Oryctolagus cuniculus* | Rabbit | Rabbit coronavirus HKU14 |
| QMT97936.1 | *Mesocricetus auratus* | Golden hamster | SARSr-CoV |
| AVN89334.1 | *Camelus dromedarius* | Arabian camel | *Middle East respiratory syndrome–related coronavirus* |
| QIA48614.1 | *Manis javanica* | Pangolin | Pangolin-CoV |
| QLG96797.1 | *Felis catus* | Domestic cat (*Felis silvestris catus*) | SARSr-CoV |
| QLC48491.1 | *Panthera tigris* | Tiger | SARSr-CoV |
| QIT08292.1 | *Canis lupus familiaris* | Domestic dog (*Canis familiaris*) | SARSr-CoV |
| ACN89742.1 | *Mus musculus* | Mouse | *Murine coronavirus* |
| AUF40275.1 | *Sus scrofa* | Wild boar | *Betacoronavirus 1* |
| BAT33329.1 | *Sus scrofa domesticus* | Domestic pig | *Porcine epidemic diarrhea virus* |
| QII89061.1 | *Tursiops truncatus* | Atlantic bottle-nosed dolphin (*Delphinus truncatus*) | Bottlenose dolphin coronavirus |
| ABW87820.1 | *Delphinapterus leucas* | Beluga whale | *Beluga whale coronavirus SW1* |
| AY515512.1 | *Paguma larvata* | Masked palm civet | SARSr-CoV |
| AJA91207.1 | *Rattus norvegicus* | Rat | *China Rattus coronavirus HKU24* |
| QJS39579.1 | *Mustela lutreola biedermanni* | European mink | SARSr-CoV |
| ASR18938.1 | *Mustela putorius furo* | Ferret (*Mustela furo*) | Ferret coronavirus |
| ATO98145.1 | *Rhinolophus ferrumequinum* | Greater horseshoe bat | SARSr-CoV |
| QDF43810.1 | *Cynopterus sphinx* | Greater short-nosed fruit bat | Coronavirus BtRs-AlphaCoV/YN2018 |
| ACT10983.1 | *Bos taurus* | Bovine | *Betacoronavirus 1* |
| ABD75332.1 | *Rhinolophus macrotis* | Big-eared horseshoe bat | SARSr-CoV |
| ATO98205.1 | *Rhinolophus sinicus* | Chinese rufous horseshoe bat | SARSr-CoV |
| ASL24654.1 | *Myotis lucifugus* | Little brown bat | *Bat coronavirus CDPHE15* |
| AVP78031.1 | *Rhinolophus pusillus* | Least horseshoe bat | SARSr-CoV |

| AOG30822.1 | *Rousettus leschenaultii* | Leschenault's rousette | *Rousettus bat coronavirus GCCDC1* |
| AGX27810.1 | *Erinaceus europaeus* | Western European hedgehog | *Hedgehog coronavirus 1* |
| ACJ66977.1 | *Odocoileus virginianus* | White-tailed deer | *Betacoronavirus 1* |
| AIA62343.1 | *Pipistrellus abramus* | Japanese house bat | *Pipistrellus bat coronavirus HKU5* |
| ASM61973.1 | *Gallus gallus* | Red junglefowl | *Avian coronavirus* |
| ACV87276.1 | *Meleagris gallopavo* | Wild turkey | *Avian coronavirus* |

# Results

**Animal susceptibility**

To investigate the correlation between animal susceptibility and ACE2 similarity, a phylogenetic analysis of human and animal ACE2 sequences is performed. A total of 225 sequences (130 unique species) obtained from UniProt and NCBI databases are included in the analysis. From the phylogenetic tree file, the similarity of each sequence to hACE2 is calculated by the distance of each branch between human and animal nodes. Comparing known animal susceptibility to ACE2 phylogenetic distances reveals that all SARS-CoV-2 susceptible animals known to date[20–27] have ACE2 distances below 0.41 (Table 3). The non-susceptible *Gallus gallus* (red junglefowl) has an ACE2 distance of 0.94 and is significantly further away from humans and susceptible animals. The only exception is *Sus scrofa domesticus* (domestic pig), whose ACE2 distance is within the range of all susceptible animals, but no viral shedding or ribonucleic acid (RNA) is detected. The ACE2 distances observed in animals can be divided into three groups: animals with an ACE2 distance within 0.5 in humans, animals with an ACE2 distance above 0.5 and below 2, and finally animals with an ACE2 distance above 2 (Figure 2). Our analysis of the ACE2 phylogenetic distance with reference to hACE2 reveals that the susceptible animals identified to date belong to the first group closest to hACE2.

**Table 3. ACE2 phylogenetic distance is lower in susceptible animals.** To date, most animals known to be susceptible have been experimentally infected. Naturally infected animals include tiger, European mink, and white-tailed deer. Susceptible animals have an ACE2 distance of 0.41 or less.

| Animal | Common name | Clinical presentation | Susceptibility | Phylogenetic distance |
|---|---|---|---|---|
| *Macaca mulatta* | Rhesus macaque | 5 days post-infection (dpi), mild to moderate symptoms[24]. | Yes | 0.06 |
| *Oryctolagus cuniculus* | European rabbit | $10^6$ fifty-percent tissue culture infective doses ($TCID_{50}$). No clinical symptoms. Viral RNA peak 2 dpi nose/throat[22]. | Yes | 0.22 |
| *Mesocricetus auratus* | Golden hamster | $10^5$ plaque-forming units (PFU), 100μL. Viral RNA detected at 2 dpi. Transmission presents[25]. | Yes | 0.25 |
| *Felis catus* | Domestic cat | $10^5$ PFU. Viral RNA detected in subadult cat | Yes | 0.27 |
| | | feces at 3 dpi. Viral RNA detected in juvenile cat nasal wash at 2 dpi[20]. | | |
| *Panthera tigris* | Tiger | Natural infection[27]. | Yes | 0.28 |
| *Canis lupus familiaris* | Domestic dog | $10^5$ PFU. Viral RNA detected in feces at 2 dpi. Negative nasal swab[20]. | Yes | 0.28 |
| *Sus scrofa domesticus* | Domestic pig | Not susceptible[20] | No | 0.29 |
| *Tupaia belangeri chinensis* | Chinese treeshrew | $10^6$ PFU, 1mL. Viral RNA detected at 6 dpi (first sample), shedding[23]. | Yes | 0.31 |
| *Mustela lutreola biedermanni* | European mink | Natural infection[3]. | Yes | 0.31 |
| *Mustela putorius furo* | Ferret | 1. $10^5$ TCID$_{50}$. Shedding detected in nasal wash at 2 dpi. Peak viral RNA nasal wash at 2 dpi[20]. 2. $10^5$ PFU. Peak viral RNA nasal wash at 4 dpi, limited replication in other organs[21]. | Yes | 0.32 |
| *Odocoileus virginianus* | White-tailed deer | Natural infection[26]. | Yes | 0.41 |
| *Gallus gallus* | Red junglefowl | Not susceptible[20,21] | No | 0.94 |

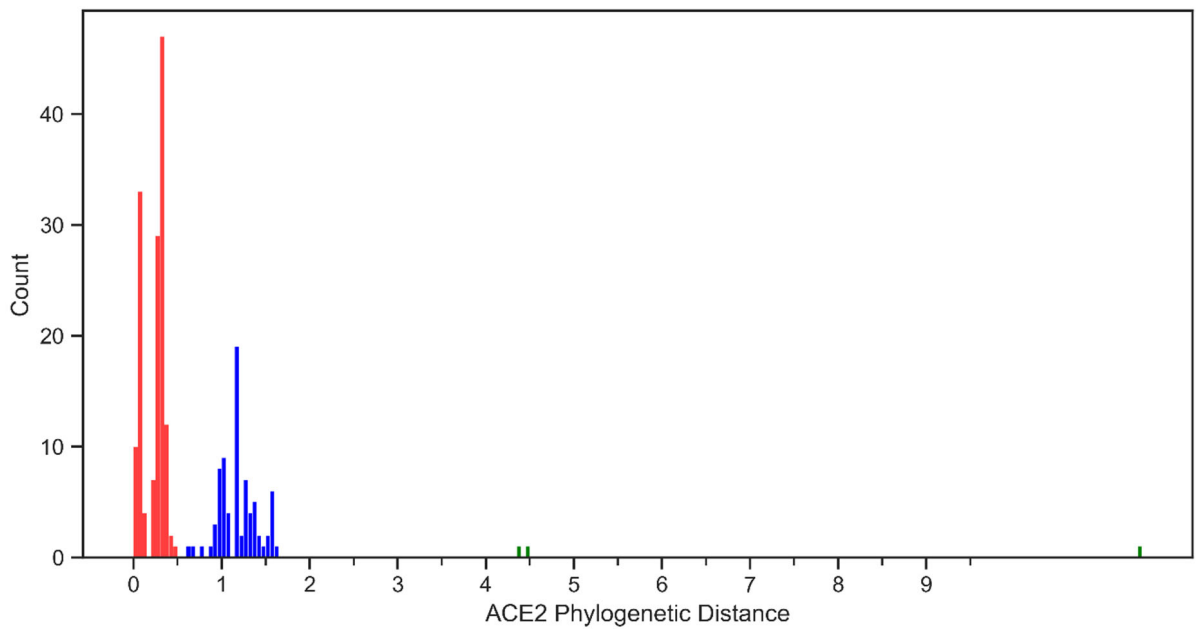

**Figure 2. Distribution of animal ACE2 phylogenetic distances.** In most animals, the branch distance between ACE2 and hACE2 is less than 1.5. Three distinct groups of distribution can be observed: animal ACE2 close to humans (red < 0.5), distant from humans (0.5 ≤ blue ≤ 2), and outliers (green). The third most distant group of animals belonging to the order Insecta is *Melanaphis sacchari* (sugarcane aphid), *Schizaphis graminum* (greenbug aphid), and *Bactrocera latifrons* (Malaysian fruit fly) (*Chaetodacus latifrons*).

**Predicting animals at risk of human spillover**

To identify potential animals that may serve as amplification hosts for SARS-CoV-2, the role of phylogenetic distance of the coronavirus spike protein in cross-species jumping is further investigated by analysing together with the ACE2 distance (Figure 3). Phylogenetic analysis of the complete sequence is studied to trace the evolutionary history of SARS-CoV-2[28,29]; however, in order to understand short-term cross-species transmission rather than long-term evolutionary history[30], only the spike protein is analysed as it is the main determinant of susceptibility. The 2105 complete spike protein sequences of coronaviruses used in this analysis are obtained from the NCBI virus database[31]. ACE2 in each animal is associated with the spike protein of the coronavirus found in that particular animal, which is known as a parasitic coronavirus.

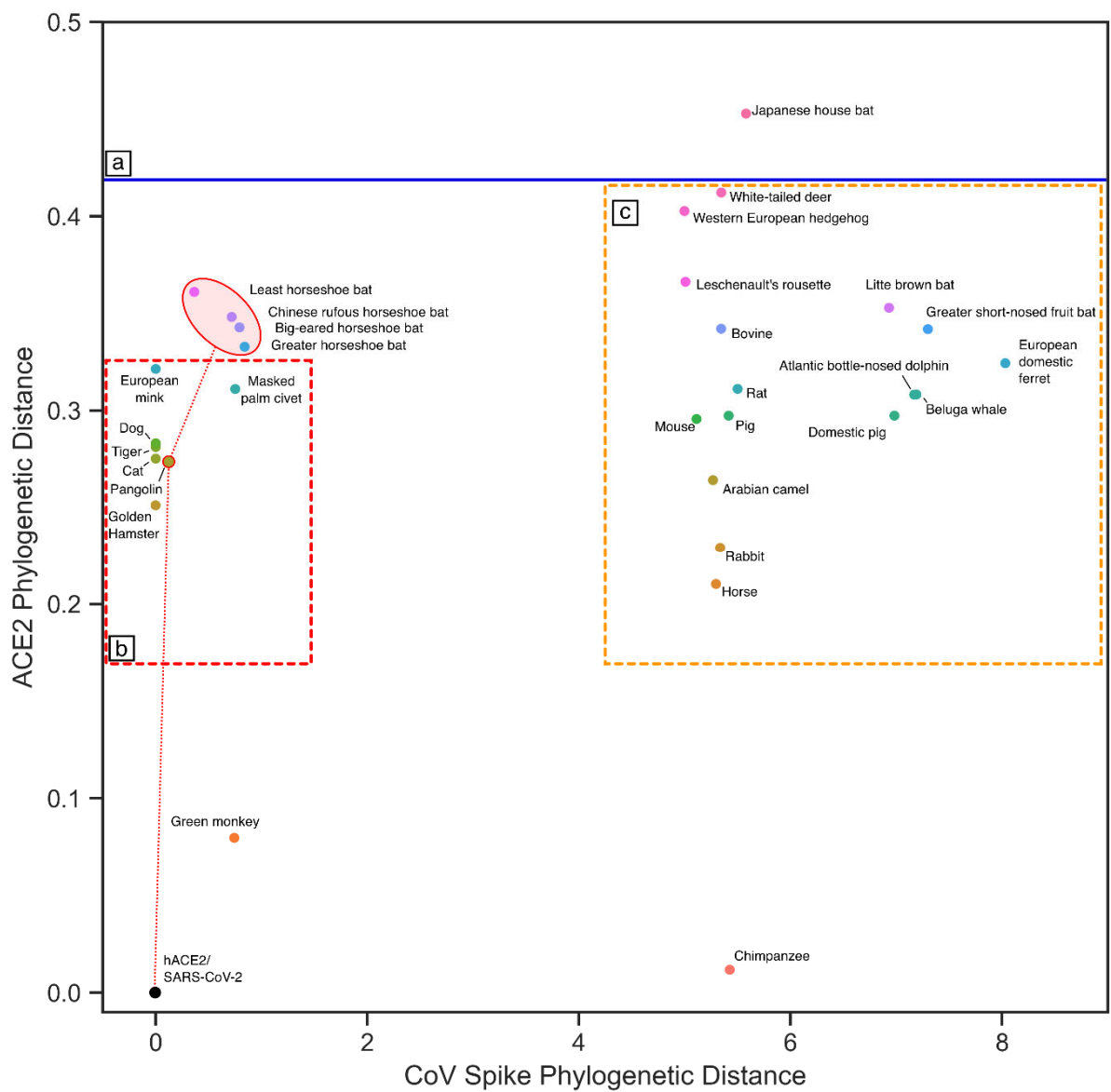

**Figure 3. Phylogenetic distance between animal ACE2 and the respective parasitic coronavirus spike protein with reference to hACE2 and SARS-CoV-2.** Only animals with ACE2 below 0.5 (group closest to hACE2) are shown. Potential amplification hosts of SARS-CoV-2: a, animals below the susceptibility line have closer ACE2 similarity to humans. This zone is categorised into two areas: b, elevated risk of SARS-CoV-2 mutation amplification (red box) and c, potential risk (orange box). High-risk animals are susceptible based on the ACE2 similarity and have a parasitic coronavirus spike protein similar to SARS-CoV-2. ACE2, which belongs to animals with unconfirmed potential, is to humans; however, there is no parasitic coronavirus spike protein similar to SARS-CoV-2.

The SARS-CoV-2 reference sequence Wuhan-Hu-1 (NC_045512.2) and spike protein (YP_009724390.1) were used as a reference to identify potential intermediate hosts or amplification hosts of the COVID-19 pandemic (Figure 3). Pangolins are suspected to be the intermediate host[32,33], and their ACE2 shares similarities with *Rhinolophus* bats and hACE2. As with the SARS-CoV transmission pathway, this jump increases the similarity between the ACE2 and spike protein sequences compared to hACE2 and the first SARS-CoV-2 spike protein sequence. A closer similarity to Pangolin-CoV suggests that SARS-CoV-2 adopts the spike protein of Pangolin-CoV, allowing recombination to occur. Animals confirmed to be susceptible (Table 3) have an ACE2 distance of 0.41 and below. The ACE2 distance of these susceptible animals belongs to the first group (ACE2 distance < 0.5). This is consistent with the intermediate animals of SARS-CoV and suspected SARS-CoV-2, namely masked palm civets and pangolins, whose ACE2 distances are 0.30 and 0.27, respectively. Thus, it is speculated that animals with an ACE2 distance less than 0.5 may be susceptible to SARS-CoV-2 infection (Table 4). Animals that have been identified as susceptible (Figure 3b) are at elevated risk of playing a role in a spillover incident. In addition to this, animals that have not been identified as susceptible (Figure 3c) should not be ignored. Since their ACE2 is closer to humans, they also at risk of infection. The nature of this analysis depends on the range of zoonotic coronaviruses that have been sequenced and made available in the database. Many animals belonging to this group may carry coronaviruses similar to SARS-CoV-2 but have not yet been identified.

**Table 4. Animals may be susceptible to SARS-CoV-2** based on an ACE2 distance between 0.5 and humans.

| Scientific Name | Common Name | ACE2 Distance |
|---|---|---|
| *Pan paniscus* | Pygmy chimpanzee | 0.012 |
| *Pan troglodytes* | Chimpanzee | 0.012 |
| *Gorilla gorilla gorilla* | Western lowland gorilla | 0.012 |
| *Pongo abelii* | Sumatran orangutan | 0.023 |
| *Nomascus leucogenys* | Northern white-cheeked gibbon | 0.028 |
| *Rhinopithecus roxellana* | Golden snub-nosed monkey | 0.059 |
| *Macaca mulatta* | Rhesus macaque | 0.059 |
| *Macaca nemestrina* | Pig-tailed macaque | 0.059 |
| *Papio anubis* | Olive baboon | 0.060 |
| *Macaca fascicularis* | Crab-eating macaque | 0.061 |
| *Colobus angolensis palliatus* | Peters' Angolan colobus | 0.063 |
| *Cercocebus atys* | Sooty mangabey | 0.065 |
| *Mandrillus leucophaeus* | Drill | 0.066 |
| *Chlorocebus sabaeus* | Green monkey | 0.074 |
| *Rhinopithecus bieti* | Black snub-nosed monkey | 0.090 |
| *Cebus capucinus imitator* | Panamanian white-faced capuchin | 0.101 |
| *Aotus nancymaae* | Ma's night monkey | 0.109 |
| *Saimiri boliviensis boliviensis* | Bolivian squirrel monkey | 0.117 |
| *Callithrix jacchus* | White-tufted-ear marmoset | 0.119 |
| *Equus caballus* | Horse | 0.213 |
| *Heterocephalus glaber* | Naked mole-rat | 0.229 |
| *Propithecus coquereli* | Coquerel's sifaka | 0.231 |
| *Oryctolagus cuniculus* | European rabbit | 0.232 |
| *Chinchilla lanigera* | Long-tailed chinchilla | 0.238 |
| *Tarsius syrichta* | Philippine tarsier | 0.238 |

| *Ictidomys tridecemlineatus* | Thirteen-lined ground squirrel | 0.242 |
|---|---|---|
| *Mesocricetus auratus* | Golden hamster | 0.254 |
| *Camelus dromedarius* | Arabian camel | 0.267 |
| *Manis pentadactyla* | Chinese pangolin | 0.274 |
| *Manis javanica* | Pangolin | 0.276 |
| *Felis catus* | Domestic cat | 0.277 |
| *Lynx canadensis* | Canada lynx | 0.278 |
| *Panthera tigris* | Tiger | 0.283 |
| *Canis lupus familiaris* | Domestic dog | 0.285 |
| *Nyctereutes procyonoides* | Raccoon dog | 0.288 |
| *Otolemur garnettii* | Small-eared galago | 0.290 |
| *Ursus americanus* | American black bear | 0.292 |
| *Dipodomys ordii* | Ord's kangaroo rat | 0.292 |
| *Phodopus campbelli* | Campbell's dwarf hamster | 0.293 |
| *Ursus arctos horribilis* | Grizzly bear | 0.293 |
| *Vulpes vulpes* | Red fox | 0.294 |
| *Ursus maritimus* | Polar bear | 0.296 |
| *Callorhinus ursinus* | Northern fur seal | 0.296 |
| *Mus musculus* | Mouse | 0.298 |
| *Physeter macrocephalus* | Sperm whale | 0.299 |
| *Lipotes vexillifer* | Yangtze river dolphin | 0.300 |
| *Ailuropoda melanoleuca* | Giant panda | 0.301 |
| *Odobenus rosmarus divergens* | Pacific walrus | 0.306 |
| *Neophocaena asiaeorientalis* | Yangtze finless porpoise | 0.306 |
| *Neomonachus schauinslandi* | Hawaiian monk seal | 0.307 |
| *Balaenoptera acutorostrata scammoni* | North Pacific minke whale | 0.310 |
| *Procyon lotor* | Raccoon | 0.310 |
| *Tursiops truncatus* | Atlantic bottle-nosed dolphin | 0.311 |
| *Delphinapterus leucas* | Beluga whale | 0.311 |
| *Paguma larvata* | Masked palm civet | 0.313 |
| *Rattus norvegicus* | Rat | 0.314 |
| *Tupaia glis* | Common treeshrew | 0.315 |
| *Tupaia belangeri chinensis* | Chinese treeshrew | 0.318 |
| *Mustela erminea* | Short-tailed weasel | 0.321 |
| *Suricata suricatta* | Meerkat | 0.321 |
| *Mustela lutreola biedermanni* | European mink | 0.324 |
| *Megaderma lyra* | Greater false vampire bat | 0.325 |
| *Mustela putorius furo* | Ferret | 0.327 |
| *Rhinolophus ferrumequinum* | Greater horseshoe bat | 0.336 |
| *Tadarida brasiliensis* | Mexican free-tailed bat | 0.338 |
| *Melogale moschata* | Chinese ferret-badger | 0.338 |
| *Ovis aries* | Sheep | 0.339 |
| *Capra hircus* | Goat | 0.340 |
| *Bos indicus x Bos taurus* | Hybrid cattle | 0.344 |
| *Bos taurus* | Bovine | 0.345 |

| *Rhinolophus macrotis* | Big-eared horseshoe bat | 0.346 |
| *Pteropus vampyrus* | Large flying fox | 0.346 |
| *Loxodonta africana* | African bush elephant | 0.349 |
| *Rhinolophus sinicus* | Chinese rufous horseshoe bat | 0.352 |
| *Rhinolophus pearsonii* | Pearson's horseshoe bat | 0.353 |
| *Arctonyx collaris* | Hog badger | 0.353 |
| *Cynopterus sphinx* | Greater short-nosed fruit bat | 0.355 |
| *Myotis lucifugus* | Little brown bat | 0.357 |
| *Rhinolophus pusillus* | Least horseshoe bat | 0.365 |
| *Rousettus leschenaultii* | Leschenault's rousette | 0.370 |
| *Cavia porcellus* | Guinea pig | 0.393 |
| *Erinaceus europaeus* | Western European hedgehog | 0.407 |
| *Odocoileus virginianus* | White-tailed deer | 0.408 |
| *Pipistrellus abramus* | Japanese house bat | 0.457 |

In April 2020, SARS-CoV-2 infection was detected in farmed mink in the Netherlands[3,34]. Mink-related mutations found in humans confirm spillover incidents from humans to mink[35,36]. In samples from feral cats near ten farms, antibodies and viral RNA were detected in the cats[3,34]. As shown in Figure 4, SARS-CoV-2 has the potential to be transmitted into farmed minks through workers (Figure 4a). Due to the close proximity of farmed minks and wandering stray cats, these animals were co-infected (Figure 4b). When workers came into contact with farmed minks, the mutated variants were spilled back into the human cycle. In summary, our analysis of the phylogenetic distance between ACE2 and the parasitic coronavirus spike protein provides a key indicator of animals that may be susceptible to SARS-CoV-2 infection and, most importantly, those animals at elevated risk of spillover and spillback.

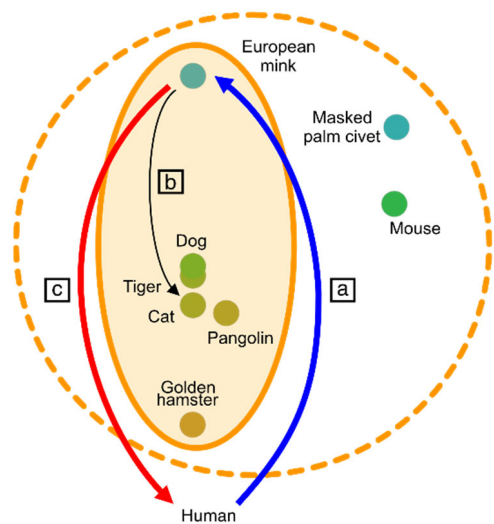

**Figure 4. Amplification cycle of minks and cats in Netherlands.** a, Spillover incident of farmed minks or stray cats. b, Farmed minks and stray cats are co-infected due to being too close together. c, The spillback incident occurred when farm workers were managing minks nearby.

# Discussion

The cellular receptor ACE2 is critical for SARS-CoV-2 binding to cells, but only after the spike protein is cleaved by the transmembrane protease, serine 2 (TMPRSS2) and furin[37–39]. Hence, the location and abundance of ACE2 expression determines the tropism for SARS-CoV-2 infection, as well as the presence of TMPRSS2 and furin, which are major determinants of whether an animal is susceptible to infection. A study comparing specific binding residues in hACE2 showed that pig ACE2 differs from humans at specific sites, explaining why pigs are resistant to infection[40]. Nevertheless, all animals confirmed to be susceptible to date exhibit a similar characteristic; namely, ACE2 phylogenetic distance below 0.5.

Two components are provided to determine the potential of an animal to serve as an amplification host: the distance between ACE2 and hACE2 as well as the distance between the parasitic protein and the SARS-CoV-2 spike protein. The path of transmission to humans starts with the bat, which is the natural reservoir, and then jumps downward to an intermediate animal that is to the left of both SARS-CoV and SARS-CoV-2. This suggests that potential intermediate animals have two characteristics. First, its ACE2 is more similar to hACE2, thereby reducing the cross-species barrier. Second, the parasitic coronavirus spike protein shares more similarities with SARS-CoV-2. This enables the platform for recombination to occur, which, coupled with the smaller ACE2 gap, could lead to the creation of a new type of coronavirus. To date, there are two known large-scale human-to-animal spillover events: the mink farm in the Netherlands and the white-tailed deer in the United States. Both animals are in close contact with humans; however, only the mink farm incident resulted in a spillback event of SARS-CoV-2 back to 68% of farm workers[35], whereas white-tailed deer had 3% (3/92) event of spillback to humans[26]. Judging from the results in Figure 3, European mink is closer to the parasitic coronavirus spike distance of SARS-CoV-2 than white-tailed deer. This suggests that animals in close proximity to ACE2 and the parasitic coronavirus spike distance have a higher risk of spillover from humans and spillback to humans and can readily serve as amplification hosts.

Besides the intermediate host, there are multiple animals with the same characteristics: ACE2, which is similar to humans, and the parasitic coronavirus spike protein, which is similar to SARS-CoV-2. This raises a similar question to the previous SARS outbreak: Are there multiple intermediate hosts[41,42]? Two other animals, the raccoon dog (*Nyctereutes procyonoides*) and the Chinese ferret-badger (*Melogale moschata*), have also been found to be infected with SARS-CoV[41]. These findings encourage further studies to determine whether multiple animals played a role in the outbreak of the novel coronavirus in 2019. Nevertheless, regardless of whether they played a role in the outbreak of the novel coronavirus in 2019, these animals could serve as amplification hosts in a spillover incident. As shown in Figure 4, these animals have been identified as susceptible and can be used as a guide for monitoring potential spillover incidents, particularly for domestic farmed animals. The recent mink spillover and spillback incident did not raise new strain of concern; however, further spillover and spillback incidents could lead to the emergence of new SARS-CoV-2 strains. Such an oversight could render SARS-CoV-2 vaccines ineffective.

# Conclusion

Our *in-silico* analysis is not a substitute for *in vitro* and *in vivo* studies of animal susceptibility or serology to confirm intermediate animals prior to an outbreak; however, it is a rapid way to identify zoonotic species that may be responsible for an outbreak or involved in spillover and spillback events. Since animal acquisition and testing is a daunting task, this approach can rapidly identify animals at high-risk to prioritise research and assess the risk of zoonotic amplification in the environment. These data were obtained from coronaviruses discovered in zoonotic hosts and in the ACE2 databases prior to the SARS-CoV-2 outbreak, except for the first reference to the Wuhan SARS-CoV-2 spike sequence. Without the aid of experimental studies, the method identifies a variety of animals that are subsequently shown to be susceptible to SARS-CoV-2 infection. This highlights the advantages of leveraging the vast amount of sequence data in the databases as a rapid response tool in the initial stages of a pandemic.

## Abbreviations

SARS-CoV-2: Severe acute respiratory syndrome coronavirus 2; SARS: Severe acute respiratory syndrome; SARS-CoV: Severe acute respiratory syndrome coronavirus; hACE2: Human angiotensin-converting enzyme 2; ACE2: Angiotensin-converting enzyme 2; Pangolin-CoV: SARS-CoV-2-related pangolin coronavirus; NCBI: National Center for Biotechnology Information; CDS: Coding sequence; SARSr-CoV: *Severe acute respiratory syndrome–related coronavirus*; RNA: Ribonucleic acid; dpi: Days post-infection; TCID: Tissue culture infective doses; PFU: Plaque-forming units; TMPRSS2: Transmembrane protease, serine 2.

## Availability of data and material

All data are available from the NCBI virus database. The Nextflow code is accessible on GitHub link as follows: https://github.com/zihian/spillover_back_model.

## Acknowledgments

This research is supported by the Ministry of Education, Singapore, under its Academic Research Fund Tier 1 (RG75/20).

Source: *Journal of Microbiology Immunology and Infection*, Vol. 57, No. 2, pp. 225-237, 2024;
DOI: 10.1016/j.jmii.2024.01.002# References

1. Zhu N, Zhang DY, Wang WL, Li XW, Yang B, Song JD, et al. A novel coronavirus from patients with pneumonia in China, 2019. *New Engl J Med* 2020;**382**(8):727–733.
2. Wu F, Zhao S, Yu B, Chen Y-M, Wang W, Song Z-G, et al. A new coronavirus associated with human respiratory disease in China. *Nature* 2020;**579**(7798):265–269.
3. Oreshkova N, Molenaar RJ, Vreman S, Harders F, Munnink BBO, Hakze-van der Honing RW, et al. SARS-CoV-2 infection in farmed minks, the Netherlands, April and May 2020. *Eurosurveillance* 2020;**25**(23):2001005.
4. Wang QH, Zhang YF, Wu LL, Niu S, Song CL, Zhang ZY, et al. Structural and functional basis of SARS-CoV-2 entry by using human ACE2. *Cell* 2020;**181**(4):894–904.
5. Shang J, Ye G, Shi K, Wan YS, Luo CM, Aihara H, et al. Structural basis of receptor recognition by SARS-CoV-2. *Nature* 2020;**581**(7807):221–224.
6. Zhao XS, Chen DY, Szabla R, Zheng M, Li GL, Du PC, et al. Broad and differential animal angiotensin-converting enzyme 2 receptor usage by SARS-CoV-2. *J Virol* 2020;**94**(18):e00940-20.
7. Johansen MD, Irving A, Montagutelli X, Tate MD, Rudloff I, Nold MF, et al. Animal and translational models of SARS-CoV-2 infection and COVID-19. *Mucosal Immunol* 2020;**13**(6):877–891.
8. Edgar RC. MUSCLE: multiple sequence alignment with high accuracy and high throughput. *Nucleic Acids Res* 2004;**32**(5):1792–1797.
9. Shu J-J, Ouw LS. Pairwise alignment of the DNA sequence using hypercomplex number representation. *B Math Biol* 2004;**66**(5):1423–1438.
10. Shu J-J, Li Y. Hypercomplex cross-correlation of DNA sequences. *J Biol Syst* 2010;**18**(4):711–725.
11. Shu J-J, Yong KY, Chan WK. An improved scoring matrix for multiple sequence alignment. *Math Probl Eng* 2012;**2012**(490649):1–9.
12. Shu J-J, Yong KY. Identifying DNA motifs based on match and mismatch alignment information. *J Math Chem* 2013;**51**(7):1720–1728.
13. Shu J-J. A new integrated symmetrical table for genetic codes. *BioSystems* 2017;**151**:21–26.
14. Di Tommaso P, Chatzou M, Floden EW, Prieto Barja P, Palumbo E, Notredame C. Nextflow enables reproducible computational workflows. *Nat Biotechnol* 2017;**35**(4):316–319.
15. Katoh K, Standley DM. MAFFT multiple sequence alignment software version 7: improvements in performance and usability. *Mol Biol Evol* 2013;**30**(4):772–780.
16. Minh BQ, Schmidt HA, Chernomor O, Schrempf D, Woodhams MD, von Haeseler A, et al. IQ-TREE 2: new models and efficient methods for phylogenetic inference in the genomic era. *Mol Biol Evol* 2020;**37**(5):1530–1534.19